\documentstyle[prl,aps,epsf, amsfonts]{revtex}

\def\8{\infty}

\def\undertext#1{\vtop{\hbox{#1}\kern 1pt \hrule}}

\def\VEV#1{\left\langle\,#1\,\right\rangle}

\def\pp#1{\frac{\partial}{\partial#1}}

\def\br{\\ \nonumber & &}

\def\be{\begin{equation}}
\def\ee{\end{equation}}
\def\bea{\begin{eqnarray} & &}
\def\eea{\end{eqnarray}}
\def\ct#1{\cite{#1}}
\def\rf#1{(\ref{#1})}

\def\ZZ{Z\!\!\!Z}

\title{Parafermion Statistics and Quasihole Excitations for the 
Generalizations of the Paired Quantum Hall States}

\author {V. Gurarie$^a$, E. Rezayi$^{a,b}$ }

\address{$^a$ Institute for Theoretical Physics, University of California,
Santa Barbara CA 93106-4030}

\address{$^b$ Department of Physics, California State University, Los Angeles, 
CA 90032}

\date{\today}

\begin {document}
\draft
\maketitle

\begin{abstract}
We continue the program started in \ct{RR} and explain the 
statistics of the excitations for the generalizations of the paired states
in the quantum Hall effect
in terms of the parafermion statistics. We show
that these excitations behave as combinations of bosons and
parafermions. That generalizes the prior treatment of the paired
(Pfaffian) state where the excitations 
behave as combinations of bosons and fermions. We explain what
it means, from a quantum mechanical point of view, for a particle to
be a `parafermion' rather than a boson or a fermion and work through
several explicit examples. 
The resulting multiplets
coincide exactly with the angular momentum
multiplets found numerically for
the $k+1$ particle interaction Hamiltonian on a sphere. 
We also present a proof that the wave functions found in \ct{RR} are indeed
the correlation functions of the parafermion conformal field theory.

\end{abstract}

\vspace{.5in}

\section {Introduction}

The quantum Hall plateaus on the first excited Landau levels appear to be
different from those of the lowest Landau level. The most
striking difference is the $5/2$ plateau \ct{Willet} for
which there is no analog in the lowest Landau level. 

To explain the $5/2$ plateau, several trial wave functions were proposed. 
One of them, the Haldane-Rezayi state \ct{HR}, assumes that the
electrons in it are spin unpolarized. Yet another trial wave function,
the Pfaffian \ct{RM}, assumes that electrons are polarized, just as for
the ordinary Laughlin trial wave functions \ct{Laughlin}. Recent numerical
evidence \ct{morf,fdmh}, in the absence of the Landau level mixing, 
supports the Pfaffian as a trial wave function for the 
$5/2$ plateau. 

In the paper by N. Read and one of us 
\ct{RR} a further generalization of the Pfaffian state was
proposed. Moreover, it was conjectured, and supported by the numerical
evidence, that these generalizations 
may be better candidates for other plateaus
on the first excited Landau levels than the 
Laughlin state and its hierarchical 
\ct{Laughlin,Haldane,Halperin} or 
composite fermion \ct{Jain} generalizations. 


An unusual feature of the generalizations of the Pfaffian state 
given in \ct{RR}
was that
the trial wave functions were found explicitly for nonsimple fractions. 
That was done with the help of conformal field theory.

Conformal field theory discovered in \ct{BPZ}
is essentially a method to solve various 
scale invariant 1+1 dimensional quantum field theories exactly.
It was since proved extremely useful for understanding various
1 dimensional quantum and 2 dimensional classical statistical mechanics
systems.
Its relevance for the quantum Hall systems was first discussed in \ct{RM}. 

The relevant conformal field theory for the generalization
of the Pfaffian is the so-called parafermion conformal
field theory  \ct{ZF}. The parafermions, the direct
generalizations of the fermions, play an important
role in describing the critical points of the $Z_k$ invariant statistical
mechanics 
systems. A particularly well known example of the $Z_2$ invariant
system is the Ising model. The Ising model can be described by a Majorana
fermion. A $Z_k$ invariant system must be described, on the other hand, by
$Z_k$-parafermions. 

A Majorana fermion $\psi(z)$ 
can be used to generate the Pfaffian trial wave function,
as was discussed in \ct{RM}. 
\be
\Psi_{\rm Pfaffian} = \langle \psi(z_1) \psi(z_2) \dots \psi(z_N) \rangle
\prod_{i<j\le N} (z_i - z_j)^2
\ee
The filling factor of this trial wave function is $1/2$. Moreover, the
so-called spin fields of the fermion conformal field theory (the order
parameter of the Ising model) $\sigma$ can be used to generate the
excitations of such a system,
\be
\Psi_{\rm Pfaffian}^{\rm excited} (\eta_1, \eta_2, \dots, \eta_{2 L})=
\langle \sigma(\eta_1), \sigma(\eta_2), \dots, \sigma(\eta_{2L}),
\psi(z_1), \psi(z_2), \dots,
\psi(z_N) \rangle \prod_{i,j} (\eta_i-z_j)^{1 \over 2} \prod_{i<j}
(z_i-z_j)^2.
\ee

On the other hand, 
as was discussed in \ct{ZF}, 
for each $k$ there are $k-1$ parafermions, denoted
$\psi_l$, with dimensions $h_k=l (k-l)/k$. In \ct{RR} it was proposed to
use the first of them to generate a parafermion wave function,
\be
\label{www}
\Psi_{\rm para}^M = \langle 
\psi_1(z_1) \psi_1(z_2) \dots \psi_1(z_N) \rangle
\prod_{i<j\le N} (z_i - z_j)^{M+{2 \over k}}
\ee
where $M$ has to be taken odd or even integer 
depending on whether the particles are fermions or bosons. 
The filling factor of this wave function is given by 
$\nu=k/(k M +2)$. 

It was further shown in \ct{RR} that this wave function should
be an exact ground state of a system of $N$ bosons in a magnetic
field in the presence of the $k$-body interaction Hamiltonian
\be
\label{Hamel}
H= V \sum_{i_1< i_2 < \dots < i_{k+1}} \delta^2 (z_{i_1}-z_{i_2})
\delta^2 (z_{i_2}-z_{i_3}) \dots \delta^2 (z_{i_k}-z_{i_{k+1}}),
\ee
thereby generalizing the Pfaffian case for which the interaction Hamiltonian
was \rf{Hamel} at $k=2$. We do not need to consider fermions because
their wave function can be obtained from that of bosons by a simple
Jastrow factor. 

To find the wave function directly from
\rf{www} is a difficult task.  However,
it was shown in \ct{RR} that the following explicit construction has the
correct properties (it is symmetric and vanishes whenever $k+1$ particles
coincide),
which  we reproduce for future reference. 

To write down $\Psi_{\rm para}^{(0)}$ we need to break the coordinates
of $N$ electrons into clusters of $k$ ($N$ should be divisible by $k$). 
For each pair of distinct clusters, say $z_1, \dots, z_k$ and $z_{k+1},
\dots, z_{2 k}$, we define factors $\chi$ by
\be
\label{chi}
\chi_{1,2}(z_1, \dots, z_k; z_{k+1}, \dots, z_{2 k}) =
(z_1-z_{k+1})(z_1-z_{k+2}) (z_2-z_{k+2})(z_2-z_{k+3}) \dots
(z_k-z_{2k})(z_{2k}-z_{k+1}).
\ee
The subscript $1, 2$ of $\chi_{1,2}$ labels the clusters (the first, starting
with $z_1$ and the second, starting with $z_k$). 

The wave function $\Psi_{\rm para}^{(0)}$ is defined in terms of these as
\be
\label{wave}
\Psi_{\rm para}^{(0)} = \sum_{P} \prod_{0\le r<s< N/k} \chi_{r+1, s+1}
(z_{P(kr+1)}, \dots, z_{P(k(r+1))}; z_{P(k s +1)}, \dots, z_{P(k(s+1))}).
\ee
It is not hard to see that for $k=2$ this wave function reproduces the
Pfaffian for bosons at $\nu=1$.

It was also observed in \ct{RR} that the quasihole excitations
above \rf{wave} can be described by the insertion of the spin fields
$\sigma_1$ of the parafermion conformal field theory. 
While some of the excitations for the Pfaffian state were found
explicitly in \ct{NW}, 
no explicit wave
functions for the excitations of the parafermion states 
have yet been found. 

The purpose of this paper is to continue the work begun in \ct{RR}. In
particular, we 
explain  the numerically observed degeneracies of the excitations of the
parafermion states. It turns out that to reproduce the numerically
observed degeneracies we need to assume that 
these excitations obey the parafermion
statistics. 

Here we would like to emphasize that although the parafermions have
been shown in \ct{BS} 
to obey the Haldane exclusion principle \ct{Fractional}
in the statistical sense
which is a valid description for large quantities of parafermions, 
for small number of parafermions we need to know the
concrete rules which govern the
combinatorics of parafermions. These rules asymptotically approach the
exclusion statistics rules as the number of parafermions becomes large. 
These rules are described below
and by themselves do not have much to do with
Haldane exclusion statistics. 

Our finding that the excitations of the parafermion states behave as
a
combination of bosons and parafermions is a direct generalization of the
well known fact that the excitations of the Pfaffian state
behave as a combination of bosons and fermions. 

We also show that the wave function \rf{wave} is indeed given by the
parafermion correlation function \rf{www}, thus confirming the conjecture
of \ct{RR}. 

\section{Excitations of the $Z_3$-Parafermion State on a
Sphere}

In this section we are going to explain how to calculate the degeneracy
of the excitations
of the parafermion states. To do that, we will have to construct
what could be called parafermion quantum mechanics and learn how to
add angular momenta of the quantum mechanical particles obeying
parafermion statistics. 

It was shown in \ct{RR} that 
if you put $N$ particles on a sphere with a magnetic monopole in the center,
turn on $k+1$ particle interaction \rf{Hamel} and then adjust the
total flux of the magnetic field through the surface of the
sphere to be equal to 
$N_\phi=\nu^{-1} N - (M+2)$ with $\nu$ being the filling factor, 
$\nu = k/(M k+2)$ and $M$ being either odd (for fermions) or
even (for bosons) nonnegative integer, 
you discover that the electrons settle into 
one unique zero energy ground
state with the wave function
\be
\Psi_{\rm GS} = \prod_{i<j} (z_i-z_j)^M \Psi(z_1, \dots, z_N),
\ee
with $\Psi(z_1, \dots, z_N)$ given by \rf{wave}. The total angular momentum
of that state is equal to $0$. 

Now if you start increasing the flux by units of one flux quanta, you 
discover that the zero energy eigenstates of \rf{Hamel}
are degenerate. These states can all be grouped into the
angular momentum multiplets. 

The reason for this is more or less clear. By increasing the flux by 1,
we create quasiholes. There is more than one way of creating those
quasiholes which explains the degeneracy of states at a higher flux. 

As was explained in \ct{RR} the elementary quasiholes at arbitrary $k$ are
not Laughlin quasiholes. Rather, they are quasiholes which carry
flux $1/k$ and we create $k$ of them at once. The wave functions
with quasiholes can be found with the help of conformal field theory,
by inserting the fields $\sigma_1$ of the parafermion conformal field theory
\ct{ZF,RR}.
However, we are not going to be interested in the explicit wave functions.

What we want to explain in this section 
is the numerically observed degeneracy of the states with quasiholes
on a sphere and how exactly one could generate all the degeneracies. 
We will show how the correct angular momentum
 multiplets can be obtained by putting parafermions into orbitals on a 
sphere and combining their angular momenta in a way consistent with
their fractional statistics. That generalizes the prior treatment
of the Pfaffian state in \ct{RR1}.

Now we would like to present, first without explanation, the rules for
finding the degeneracies of the excitations above the $k=3$ state, that is,
for $Z_3$-parafermions.

It was found in \ct{RR}, by matching the numerical data, 
that, as we increase the flux, the degeneracy
of states we get for the $Z_3$ parafermions
is given by the following formulas. 
The number of excitations at the excess flux 1 (3 quasiholes, each
carrying $1/3$ of the  flux quantum)
is given by a binomial 
coefficient
\be
\label{firstcoef}
{ {N/3+3} \choose 3},
\ee
with $N$ being the total number of the electrons. $N$ should
be divisible by three. This coefficient has
a simple interpretation as the number of ways you can put 3 bosons 
into $N/3+1$ orbitals. The reason why the excitations at the
excess flux $1$ behave as bosons was explained in \ct{RR} and is essentially
the same for the
parafermion states as for the Pfaffian state \ct{RR1}. 

Moreover, following \ct{RR1}, we can assign the angular momentum quantum
numbers to these states. For this purpose we interpret the $N/3+1$ states
as the orbitals on a sphere, in the multiplet of the angular momentum
$L=N/6$.
When we put the 
bosons on this sphere, we generate the angular momentum multiplets
by combining their angular momenta while 
keeping their wave function totally
symmetric. 

While we could refer to any standard quantum mechanics textbooks for
the rules on how to combine the angular momenta of bosons, we
can recast these rules into the following simple form. 
Let us visualize $N/3+1$ 
orbitals as boxes. Each box has a number $L_z$ assigned to it which
varies from $-N/6$ to $N/6$ by steps of $1$. We
are allowed to put as many bosons as we wish into each particular box. 

\begin{figure}[tbp]
\centerline{\epsfxsize=2in \epsfbox{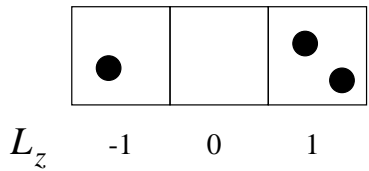}}
\caption{One possible way to put 3 bosons into 3 boxes, realizing
\rf{firstcoef} for $N=6$}
\label{Fig1}
\end{figure}

For example, for $N=6$ we have $3$ boxes with $L_z=-1$, $L_z=0$, and
$L_z=1$. 
Fig. \ref{Fig1} shows one possible way to put 3 bosons into
3 boxes, by putting 2 of them into $L_z=1$ box and one into $L_z=-1$ box,
thus giving the total $L_z=1$. 
Of course, there are ${5 \choose 3 }=10$ ways to put 3 bosons
into 5 boxes. Out of these 10 ways, the angular momentum projections
$L_z=\pm 3$ or $L_z=\pm 2$
can be obtained in 1 way each, while the angular momentum projections
$L_z=\pm 1$ or $L_z=0$ can be obtained in 2 different
ways each. We immediately recognize
the sum of the $L=3$ and $L=1$ angular momentum multiplets. 
And indeed, this is confirmed by the numerical data. 

As the flux is increased we observe many more excited states.
It was found in \ct{RR} that the number of states at the 
excess flux $2$ is given by
\be
\label{par1}
{{N/3+6} \choose 6} + 3 {{N/3+5} \choose 6} + {{N/3+4} \choose 6},
\ee
and at the three  excess flux quanta
\be
\label{par2}
{{N/3+9} \choose 9}+ 10 {{N/3+8} \choose 9} + 10 {{N/3+7} \choose 9}. 
\ee

The higher fluxes turned out to be harder to analyze as there is less
numerical data available for them. 

We observe that the number of states at these flux quanta is reminiscent 
of 
the corresponding formula for the Pfaffian which was found in \ct{RR1} to be
\be
\label{pfaf1}
\sum_{F, (-1)^F=(-1)^N, F\le N} 
{n \choose F} {{(N-F)/2+2 n} \choose {2 n}}
\ee
with $n$ being the number of the excess flux quanta. 
The extra $n \choose F$ was interpreted as the number of ways one can put
$F$ fermions into $n$ boxes. By analogy, we can write down a formula
which generalizes \rf{pfaf1} to the case of the $k=3$ parafermions,
\be
\label{paraf}
\sum_{F \ge 0, \exp \left( 2 \pi F i \over 3 \right)= 
1, F \le N } 
\left\{ {n \above 0pt F} \right\}{{(N-F)/3 + 3 n} \choose {3 n}}
\ee
where $\left\{ {n \above 0pt F} \right\}$ is the number of ways you
can put $F$ $Z_3$-parafermions into $n$ orbitals. 

What does it exactly mean, to put parafermions into orbitals?
Inspired by the parafermion mode counting derived in \ct{BS},
we present the following rules. While the rules
are relatively complicated, they are the only way known to us which
allows to obtain $\left\{ {n \above 0pt F} \right\}$ and,
at the same time, to break the configurations into angular momentum
multiplets. 

\begin{figure}[tbp]
\centerline{\epsfxsize=2in \epsfbox{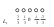}}
\caption{Positions where the parafermions are to be put for
$n=2$}
\label{Fig2}
\end{figure}

We replace the boxes by the `positions'. As we move from right to left,
each next position carries $1/3$ more of
the $z$-component of the 
angular momentum, as in Fig. \ref{Fig2}, ranging from 
$-(3 n -2)/6$ to $(3 n -2)/6$. Thus, we have
$3 n - 1$ positions. 
Fig. \ref{Fig2} depicts the positions at $n=2$. 

There are two types of the $Z_3$-parafermions we can put into these
positions. We depict  them by either the light shaded or dark shaded
circles. The following rules should be used
to put the parafermions into the positions:

\noindent
1. The light shaded circle carries $L_z$ equal to the
$L_z$ of its position.

\noindent
2. The dark shaded circle carries $L_z$ equal to twice the
$L_z$ of its position. In addition,
the dark shaded circle is counted as {\bf two} parafermions. 

\noindent
3. The dark shaded parafermion can occupy any position, while
the light shaded parafermion is not allowed to occupy the rightmost
or the leftmost position. The number of empty positions to the left
of the leftmost parafermion
has to be $3 l$ if it is dark shaded, or
$3l +1$ if it is light shaded, where $l$ is any nonnegative integer, 
including $0$. For example, if the leftmost parafermion is the light shaded
one, it can occupy the position number $2$, or  $5$, or in general
$2+3l$ counting from the 
left, and if it is a dark shaded one, it can occupy the position
number $1$, or $4$, $\dots$.

\noindent
4. The adjacent light shaded parafermions are allowed to have 
$3 l$ empty positions 
between them, while the adjacent dark shaded parafermions have to have
$3 l + 1$ empty positions between them. 
If a light shaded parafermions is adjacent to a dark shaded parafermion,
they have to be separated by $3 l + 2$ empty positions, where 
$l$ is any nonnegative
integer including $0$.

\begin{figure}[tbp]
\centerline{\epsfxsize=6in \epsfbox{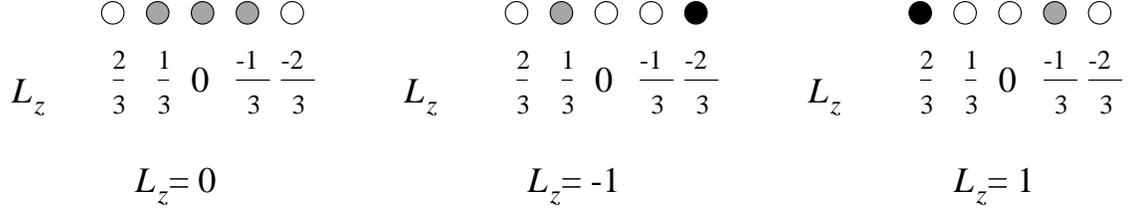}}
\caption{Three possible ways one can put 3 $Z_3$-parafermions into 
2 boxes, in accordance with $\left\{ {2 \above 0pt 3 } \right\}=3$.}
\label{Fig3}
\end{figure}

Examples of the allowed configurations are shown on Fig. \ref{Fig3}. 
This figure depicts the $3$ ways one can put $3$ parafermions into 
$5$ positions at $n=2$, consistent with \rf{par1} and \rf{paraf}. 
Moreover, we immediately observe that these configurations form an $L=1$
multiplet. 

It is not very convenient to draw positions all the time, so
we introduce the following compact notations. Each light circle will
be represented by $\psi_s$ with $s$ equal to $L_z$ of the circle.
Each dark circle will be represented by $\phi_s$. For example, the three
configurations of the Fig. \ref{Fig3} can be represented as
\be
\psi_{1\over 3} \psi_{0} \psi_{{-1 \over 3}}, \ \
\psi_{1 \over 3} \phi_{{-2 \over3} }, \ \ \phi_{2 \over 3} \psi_{{-1 \over 3}}
\ee
(we remember that in accordance with the rule 2, $\phi$ is counted as
two parafermions). 

There is only one way to put $6$ parafermions into $5$ positions at $n=2$,
\be
\phi_{2 \over 3} \phi_0 \phi_{{-2 \over 3}},
\ee
consistent with $\left\{ {2 \above 0pt 6} \right\}=1$.

Let us demonstrate the power of the technique by breaking the 
$n=3$ configurations into multiplets. There should be $8$ positions, 
carrying $L_z$ from $-7/6$ to $7/6$, that is, 
$-7/6$, $-5/6$, $-3/6$, $-1/6$, $1/6$, $3/6$, $5/6$, and $7/6$.

It is obvious there is only $1$ way to put zero parafermions into these
positions. 

According to \rf{par2} and
\rf{paraf} $\left\{ {3 \above 0pt 3 } \right\}=10$. Indeed, we
can count the configurations directly
\bea
L_z \left(\psi_{5 \over 6} \psi_{3 \over 6} \psi_{1 \over 6}\right)
= {3 \over 2}, \ \
L_z\left(\psi_{5 \over 6} \psi_{3 \over 6} \psi_{- 5 \over 6}\right)
= {1 \over 2}, \ \
L_z\left(\psi_{5 \over 6} \psi_{-3 \over 6} \psi_{-5 \over 6}\right)
= -{1 \over 2}, \ \
L_z\left(\psi_{-1 \over 6} \psi_{-3 \over 6} \psi_{- 5 \over 6} \right)
= -{3 \over 2}, \br
L_z\left(\phi_{7 \over 6} \psi_{1 \over 6}\right) 
= {5 \over 2}, \ \
L_z\left(\phi_{7 \over 6} \psi_{-5 \over 6}\right) 
= {3 \over 2}, \ \
L_z\left(\phi_{1 \over 6} \psi_{-5 \over 6}\right)
= -{1 \over 2}, \br
L_z\left(\psi_{5 \over 6} \phi_{-1 \over 6} \right) 
= {1 \over 2}, \ \
L_z\left(\psi_{5 \over 6} \phi_{-7 \over 6}\right) 
= -{3 \over 2}, \ \
L_z\left(\psi_{-1 \over 6} \phi_{-7 \over 6}\right)
= -{5 \over 2}.
\eea

We observe $10$ different configurations. Moreover, we observe
that $L_z=\pm 5/2$ occurs only once each, and $L_z=\pm 3/2$ or $L_z = 
\pm 1/2$ occurs twice each. We immediately recognize the sum of $L=5/2$ and
$L=3/2$ angular momentum multiplets. 

Quite analogously, for $6$ parafermions at $n=3$ we find
\bea
L_z \left(\psi_{5 \over 6} \psi_{3 \over 6} \psi_{1 \over 6} \psi_{-1 \over
6} \psi_{-3 \over 6} \psi_{-5 \over 6} \right)
= 0, \br
L_z\left(\psi_{5 \over 6} \psi_{3 \over 6} \psi_{1 \over 6}
\psi_{-1 \over 6} \phi_{-7 \over 6} 
\right)
= {-1}, \ \
L_z\left(\phi_{7 \over 6} \psi_{1 \over 6} \psi_{-1 \over 6}
\psi_{-3 \over 6} \psi_{-5 \over 6} 
\right)
= 1, \br
L_z\left(\phi_{7 \over 6} \phi_{3 \over 6} \psi_{- 3 \over 6} 
\psi_{-5 \over 6}
\right)
= 2, \ \ 
L_z\left(\psi_{5 \over 6} \psi_{3 \over 6}
\phi_{-3 \over 6} \phi_{-7 \over 6}
\right) 
= -2, \ \
L_z\left(\phi_{7 \over 6} \psi_{1 \over 6}
\psi_{-1 \over 6} \phi_{-7 \over 6}
\right) 
= 0, \br
L_z\left(\phi_{7 \over 6} \phi_{3 \over 6}\phi_{-1 \over 6} \right)
= 3, \ \
L_z\left(\phi_{7 \over 6} \phi_{3 \over 6}\phi_{-7 \over 6} \right)
= 1, \ \
L_z\left(\phi_{7 \over 6} \phi_{-3 \over 6}\phi_{-7 \over 6} \right)
= -1, \ \
L_z\left(\phi_{1 \over 6} \phi_{-3 \over 6}\phi_{-7 \over 6} \right)
= -3 .
\eea
The total multiplicity is $10$, in agreement with \rf{par2}. 
Counting $L_z$ we observe that $L_z=\pm 3$ and $L_z=\pm2$
occurs in one configuration each, 
while $L_z=\pm 1$ and $L_z=0$ occurs twice each. 
We recognize the direct sum of $L=3$ and $L=1$ multiplets.

One can further check that it is impossible to fit $9$ or more
parafermions into these positions.

The Tables \ref{kabs}, \ref{nidra}, and \ref{nidraban} in the 
appendix below contain the numerical  data available for the Hamiltonian
\rf{Hamel} at $k=3$ and at the number of electrons fixed at $N=6$, 
$N=9$ and $N=12$. The data is in the form $L$ versus $n$. For each excess
flux quanta $n$ the intersection of $n$th column and $L$th row contains
the multiplicity of the angular momentum $L$ at given $n$. 

To reproduce this data using the parafermion counting rules, 
we need to combine
the angular momenta of the parafermions with those of bosons, by analogy with
\rf{paraf}. This is now trivial to do, since parafermions and bosons
are distinguishable particles, and their angular momenta combine in accordance
with standard rules. 

For example, at $N=6$ we can obtain the multiplets in the following way. 
First we need to put $9$ bosons into $3$ orbitals as represented by the
term of \rf{paraf} with $F=0$. Their angular momenta are calculated 
in exactly the same way as we calculated the angular momenta for $3$ bosons
in $3$ orbitals in the text directly preceding Fig. \ref{Fig1}. 

Then we have to put $9$ bosons into $2$ orbitals and combine their angular
momenta
with the $L=5/2$ and $L=3/2$ angular momenta of the parafermions, as
represented by the term of \rf{paraf} with $F=3$. 

And finally,  we put $9$ bosons into $1$ orbital. There is only one such state
and its angular momentum is $0$. Therefore the total angular momentum
is just $L=3$ and $L=1$, coming from the parafermion contribution. 

We are not going to go through explicit counting since it is rather
standard. The only nontrivial step was the parafermion angular momentum
contribution, and that was worked out above. We have checked, however,
that the
result reproduces the numerical data given in the Table \ref{kabs}. In fact,
we have verified all the data in all three Tables, expect the
results for $n=5$ at $N=6$. In particular, we 
found that the  parafermion multiplets at $n=4$ are the following.
At $F=3$, there is one each of  $L=0,2,3,4$ states giving the
multiplicity of $22$ (compare with \rf{eq20}). At $F=6$, there are
two each of $L=0, 2, 4$ states, and one each of $L=3, 6$
states giving the overall multiplicity of $50$. For $F=9$ 
it is one of each $L=0, 2, 4$, with a total multiplicity of $15$. 
And for $F=12$ there is one $L=0$ state. 

It would be nice if we could obtain 
the parafermion multiplicities $\left\{ {n \above 0pt F } \right\}$ in a more
closed form rather than by having to count configurations all the time. 
The most closed form that
we are aware of can be obtained with the help
of the equation for the partial partition functions derived in \ct{BS}.

We define the polynomials $Y_l(x)$ in the following way. 
\be
\label{recur}
Y_{l+1}(x)=x Y_{l+ {2 \over 3}}(x) + x^2 Y_{l+{1 \over 3}}(x) +
(1 - x^3) Y_{l}(x),
\ee
with the initial conditions $Y_{- {1 \over 3}}=0$, $Y_{0}=1$, and
$Y_{1 \over 3}=0$. Then we claim that the following expansion is valid
\be
\label{defrec}
Y_n(x) = \sum_{k=0, 1, \dots} \left\{ {n \above 0pt 3 k } \right\} x^{3k}.
\ee
It is in fact possible to derive \rf{recur} directly from the counting
rules presented above, see \ct{BS}. 

It is not hard to check that
\bea
Y_1=1, \ \ Y_2 = 1 + 3 x^3 + x^6, \ \ Y_3 = 1+ 10 x^3 + 10 x^6
\eea
in agreement with \rf{par1} and \rf{par2}. 
We can continue further,
\be
\label{eq20}
Y_4=1+22 x^3 + 50x^6 + 15x^9+x^{12}, \ \ Y_5=1+ 40 x^3+168 x^6 + 140 x^9 +
28 x^{12}.
\ee
These multiplicities are also in agreement with the available numerical 
data for the degeneracies at various flux quanta $n$. 

We note that according to \ct{RR} the sum of the multiplicities at 
fixed $n$ should give us Fibonacci numbers $F_{3 n-2}$,
\be
\sum_{k=0,1,2, \dots} \left\{ {n \above 0pt 3 k } \right\} = F_{3 n-2}.
\ee

That is indeed true, since
this sum is equal to $Y_n(1)$, and by substituting $x=1$ into \rf{recur}
we obtain
\be
Y_{l+1}(1)=Y_{l+ {2 \over 3}}(1) +  Y_{l+{1 \over 3}}(1),
\ee
which is the defining relation for the Fibonacci numbers.

\section{Conformal Field Theory and the Counting Rules}

The rules we have just presented were in fact deduced from the 
parafermion conformal
field theory. Here we would like to explain how conformal field theory
should be employed to derive these rules. 

While we were interested in parafermion quantum mechanics, conformal
field theory, as a field theory, gives us the counting rules in the
second quantized formalism. It is not so hard to read 
statistics off the second quantized formalism. 

Let us concentrate on 
the Majorana
fermions as an example. 
We know, of course, that there are $n \choose F$ ways to put $F$ fermions
into $n$ boxes, in accordance with Pauli exclusion principle. This can be
obtained also from the conformal field theory of the Majorana fermions. 

Expanding the Majorana fermion $\psi_n$ in terms of modes we get
\be
\label{Majorana}
\psi(z)=\sum_{n} \psi_n z^{-n-{1 \over 2}},
\ee
with $n$ going over half integers $1/2+\ZZ$.
The modes $\psi_n$ obey the anticommutation relations,
\be
\label{commm}
\psi_n \psi_m + \psi_m \psi_n = \delta_{n+m,0}.
\ee
All these are well known facts of conformal field theory. 

Since the Majorana fermion defined at every point in space and time
cannot be infinite when acting on the
vacuum, $\psi(z)|0\rangle$ for any $z$ including $z=0$,  it follows that
$\psi_n|0\rangle=0$ for all $n=1/2,3/2,\dots$. We say that the modes of
$\psi(z)$ with positive index act as annihilation operators. The modes
$\psi_n$ with $n=-1/2, -3/2, \dots$ act as creation operators. We can
use them to create new states. A generic state will look like
\be
\label{states}
\psi_{-n_F} \psi_{-n_{F-1}} \dots \psi_{-n_1} |0 \rangle ,
\ee
with all the $n_i$ being negative  half integers. 
Note that some of the states in \rf{states} are not linearly independent
and can actually be transformed into each other by repeatedly using the
anticommutation relations \rf{commm}. We can select a linear independent
subset of \rf{states} by ordering all the $n$, say making them increase
from left to right, and making sure neither of them are equal to each other,
thereby making the state zero according to $\psi_n^2=0$. 

Now observe that there is the following correspondence between the states
\rf{states} and the angular momentum wave functions.
Consider the states  \rf{states} together with the relevant powers of the
coordinates, as in \rf{Majorana}. 
\be
\label{zcoor}
\psi_{-n_F} \psi_{-n_{F-1}} \dots \psi_{-n_1} | 0 \rangle \
 z_1^{n_F-{1 \over 2}}
z_2^{n_{F-1}-{1\over 2}} \dots z_F^{n_1- {1\over 2}}.
\ee
Let us now sum \rf{zcoor} over all the states accessible via the
anticommutation relations \rf{commm}. In other words, we want to sum
over all the permutations $\sigma$ of the numbers $n_1$, $n_2$, $\dots$,
$n_F$. Obviously we obtain
\bea
\label{mstates}
\sum_\sigma
\psi_{-\sigma(n_F)} \psi_{-\sigma(n_{F-1})} \dots \psi_{-\sigma(n_1)} 
| 0 \rangle \ z_1^{\sigma(n_F)-{1 \over 2}}
z_2^{\sigma(n_{F-1})-{1\over 2}} \dots z_F^{\sigma(n_1)- {1\over 2}} = \br
\psi_{-n_F} \psi_{-n_{F-1}} \dots \psi_{-n_1} | 0\rangle \
\sum_{\sigma} {\rm sign}(\sigma) \ z_1^{\sigma(n_F)-{1 \over 2}}
z_2^{\sigma(n_{F-1})-{1\over 2}} \dots z_F^{\sigma(n_1)- {1\over 2}}.
\eea
One immediately recognizes the totally antisymmetric polynomials
of the sort discussed in \ct{RR1}. The angular momentum $L_z$ computed
on these polynomials can be defined, up to an additive
constant,  as generated by the operator
$z {\partial \over \partial z}$, to give $\sum_i n_i- F/2$. These
totally antisymmetric polynomials are in fact the {\sl wave functions}
of the fermions in the first quantized formalism!

If we consider all possible states \rf{mstates}, with the restrictions
$n_N<n$, $n$ being some integer, there is going to be 
exactly $N \choose F$ of them which 
are the multiplicities in \rf{pfaf1}. Now we see how 
these states break naturally into
angular momentum multiplets. 
One can check that the multiplets we obtain in this way
matches the numerical data for the Pfaffian state \ct{RR1}.

What we just did so far looks like a complicated and not very intuitive
way of rederiving the results of \ct{RR1}. However, as we move on to the
parafermion states, conformal field theory becomes the only consistent
and dependable way to construct the angular momenta multiplets. 

$k=3$ parafermion conformal field theory contains two fields of dimension
$2/3$, $\psi^{(1)}(z)$ and $\psi^{(2)}(z)$ \ct{ZF}. Expanding them in terms
of modes we obtain
\be
\psi^{(1)}(z) = \sum_n 
\psi^{(1)}_n z^{-n - {2 \over 3}}, \ \ 
\psi^{(2)}(z) = \sum_n 
\psi^{(2)}_n z^{-n - {2 \over 3}}.
\ee
It is obvious that  the modes $\psi^{(1)}_n$ and $\psi^{(2)}_n$ have
to annihilate the vacuum if $n> -2/3$. Applying 
the modes with $n\le -2/3$ to the vacuum we
can in fact create parafermionic states. Not all the states we create
in this way are linearly independent. To check which ones are independent
we have to employ the generalized commutation relations of the sort derived in
\ct{ZF}. These relations are very complicated. Fortunately for $k=3$
the set of independent states have already been derived in \ct{BS}. Here
for completeness
we are going to quote the answer. 

The full set of linearly independent states is created by applying the
modes of the parafermions of the first kind $\phi^{(1)}_s
\equiv \psi^{(1)}_s$ or a certain combination of the modes
$\phi^{(2)}_s \equiv \psi^{(1)}_s \psi^{(1)}_{s - {2 \over 3}}$. 
We do not need to use
the modes of the parafermion of the second kind $\psi^{(2)}$ as its
modes create states which are linearly dependent on the states
created by $\phi^{(1)}_s$ and $\phi^{(2)}_s$.

The allowed states have the form
\be
\label{allowed}
\phi^{\left(i_N\right)}_{- s_N} \dots 
\phi^{\left(i_2\right)}_{- s_2}
\phi^{\left(i_1\right)}_{- s_1} | 0 \rangle ,
\ee
with the spacing specified as
\bea
{\rm if} \ \ i_{l+1}=1,\ i_l=1  \ \ {\rm then} \ \ 
s_{l+1}-s_l = m + {1\over 3},  \br
{\rm if} \ \ i_{l+1}=2,\ i_l=1  \ \ {\rm then} \ \
s_{l+1}-s_l = m + {2 \over 3},  \br
{\rm if} \ \ i_{l+1}=1,\ i_l=2  \ \ {\rm then} \ \
s_{l+1}-s_l = m + {4 \over 3},  \br
{\rm if} \ \ i_{l+1}=2,\ i_l=2  \ \ {\rm then} \ \
s_{l+1}-s_l = m + {2 \over 3},
\eea
where $m$ is any nonnegative integer number including $0$. 

We note that these states are in one to one correspondence with the
counting rules of the previous section. In fact, this 
is how the counting
rules should be derived. 

We can go slightly further and conjecture  a way to derive the parafermion
wave function. We need to consider the sum over all the
states which can be obtained from one of the states \rf{allowed} by the
generalized commutation relations
\be
\label{wavefu}
\sum_{\rm dependent \ states} \psi^{\left( j_N \right)}_{s_N} \dots 
\psi^{\left( j_1 \right)}_{s_1} |0 \rangle \ z_N^{- s_N- {2 \over 3}}
\dots z_1^{-s_1 - {2 \over 3}}, 
\ee
with different $j$ being either $1$ or $2$. 
By using those generalized commutation relations we can in principle bring
the sum to the form (compare with \rf{mstates})
\be
\label{defin}
\phi^{\left(i_N\right)}_{- s_N} \dots 
\phi^{\left(i_2\right)}_{- s_2}
\phi^{\left(i_1\right)}_{- s_1} | 0 \rangle \ f(z_1, \dots, z_N).
\ee
We interpret $f(z_1, \dots, z_N)$ to be the wave function of the
parafermions!  

While finding $f$ explicitly is beyond the scope of this paper,
we note that it indeed combines the single particle states of the form
$z^s$ into the multiparticle wave function obeying the right 
particle exchange properties. 


One should also in principle be able to
derive $f$ from the quasihole wave functions in direct analogy with the
corresponding derivation for the Pfaffian state \ct{NW,RR1}. This also
lies beyond the scope of this paper. 

At the end we would like to note that for $k>3$ the mode
counting rules can also be derived by using the generalized
commutation relations of \ct{ZF}. Some partial results at $k=4$ and
$k=5$ are presented in \ct{BS}.

\section {The Correlation Functions of the Parafermion CFTs}

All the preceding discussion of this paper was based on the assumption
that the wave function \rf{wave} can be generated as a correlation
function of the parafermions, in the sense of \ct{RM}. 

While it was conjectured in  \ct{RR} 
that \rf{wave} is such a correlation
function, no proof was found. It is important to prove
this relationship, otherwise, our manipulations with parafermions 
lose their relevance.

In this section
we would like to present a proof that the wave function found in \ct{RR},
which
is the zero energy state of the $k+1$ particle interaction Hamiltonian
\rf{Hamel}, is indeed the correlation function of a parafermion
conformal field theory. 
For this purpose we recall that a parafermion conformal field theory
\ct{ZF}
consists of $k-1$ fields $\psi_l$, $l=1, \dots, k-1$ with the operator
product expansion
\be
\label{OPE}
\psi_l(z) \psi_{l'}(z') \propto c_{l, l'} (z-z')^{-2 l l'/k} ( \psi_{l+l'}
+ \beta_{l, l'} (z-z') \partial \psi_{l+l'} + \dots),
\ee
where $c$ and $\beta$ are certain numbers \--- structure constants. 
The index $l$ of the fields $\psi_l$ in \rf{OPE} has to be understood in the
mod $k$ sense, $\psi_{k+l} \equiv \psi_{l}$. Additionally, $\psi_0$ is
identified with the unit operator. In that case, $\partial \psi_0$ vanishes
and therefore, when $l+l' = 0$ mod $k$, the linear term in 
\rf{OPE} proportional to $\beta$ vanishes.

According to the conjecture of \ct{RR} the correlation function
of $N$ fields $\psi_1$ is equal to
\be
\label{state}
\VEV{ \psi_1 (z_1) \dots \psi_N (z_N) } =
{ \Psi (z_1, z_2, \dots, z_N) \over \prod_{i < j} (z_i-z_j)^{2 \over k} }
\ee
with $\Psi$ given by \rf{wave}. 

To show that the right hand side of \rf{state} is indeed consistent with
\rf{OPE}, we will glue $k-1$ parafermions $\psi_1$ together
to obtain the correlation function with $\psi_{k-1}$. Then we will
glue $\psi_{k-1}$ and $\psi_1$ together and show 
that the correlation function
is consistent with the fact that
\be
\label{xxxx}
\psi_{k-1}(z) \psi_1(z') \propto (z-z')^{- 2{k-1 \over k}} \psi_0 
+ \dots
\ee 
with a vanishing first derivative $\partial \psi_0$.  

The proportionality sign in \rf{xxxx} and throughout this section
means that an unimportant numerical constant has been dropped. 

We start with taking the limit $z_1 \rightarrow z_2$. In that limit
the right hand side of \rf{state} is proportional to $(z_1-z_2)^{- 2/k}$.
This is indeed consistent with the operator product expansions
\be
\psi_1 (z) \psi_1 (z') \propto (z-z')^{-{2 \over k}} \psi_2(z') + \dots
\ee
By multiplying \rf{state} by $(z_1-z_2)^{2/k}$ and taking $z_1 \rightarrow
z_2$ we arrive at the following correlation function
\be
\label{state2}
\VEV{ \psi_2 (z_2) \psi_1 (z_3) \dots \psi_1 (z_N) }
\propto
{\Psi (z_2, z_2, z_3, \dots, z_N) \over
\prod_{i>2} (z_2-z_i)^{4 \over k} \prod_{2 < i < j \leq N}
(z_i-z_j)^{2 \over k}}.
\ee

As $z_2$ approaches $z_3$ the right hand side of \rf{state2} has just
the right singularity $(z_2-z_3)^{-4/k}$, matching the singularity
of \rf{OPE} as $\psi_2$ approaches $\psi_1$. Therefore, 
we continue this process further and take $z_2 \rightarrow z_3$.
By doing so we indeed recover the correlation function with the
field $\psi_3$,
\be
\VEV{ \psi_3 (z_3) \psi_1 (z_4) \dots \psi_1 (z_N) }
\propto
{\Psi (z_3, z_3, z_3, z_4, \dots, z_N) \over
\prod_{i>3} (z_3-z_i)^{6 \over k} \prod_{3 < i < j \leq N}
(z_i-z_j)^{2 \over k}}.
\ee

It is clear at this point that as we continue `gluing' parafermion
fields together at some point we will arrive at
\be
\label{statelast}
\VEV{ \psi_{k-1} (z_{k-1}) \psi_1 (z_k) \dots \psi_1 (z_N) }
\propto
{\Psi (z_{k-1}, z_{k-1}, \dots, z_{k-1}, z_k, z_{k+1}, \dots, z_N) \over
\prod_{i>k-1} (z_{k-1}-z_i)^{2 (k-1) \over k} \prod_{k-1 < i < j \leq N}
(z_i-z_j)^{2 \over k}}.
\ee

At this stage we should be careful. By taking $z_{k-1}$ to $z_k$ we should
be able to recover the identity operator $\psi_k \equiv \psi_0$.
Let us check that this is indeed the case. The main singularity of
\rf{statelast} as $z_{k-1} \rightarrow z_k$ is $(z_{k-1}-z_k)^{-{{k-1} \over
k}}$ which indeed matches \rf{OPE} and therefore, 
\be
\label{statenew}
\VEV{ \psi_0 (z_k) \psi_1 (z_{k+1}) \dots \psi_1 (z_N) }
\propto
{\Psi (z_k, z_k, \dots, z_k, z_{k+1}, z_{k+2}, \dots, z_N) \over
\prod_{i>k} (z_k-z_i)^2 \prod_{k < i < j \leq N}
(z_i-z_j)^{2 \over k}}.
\ee
By its definition, the identity operator does not depend on its
position, and therefore the right hand side of \rf{statenew} should
not depend on $z_k$. To see that, let us recall
that in \ct{RR} the following theorem was proved. 
The wave function $\Psi$ for $N$ particles whose first $k$
particles live at the same point $z_k$ can be expressed in terms of the
wave function for $N-k$ particles in the following way,
\be
\label{theorem}
\Psi (z_k, z_k, \dots, z_k, z_{k+1}, \dots, z_N) \propto
\prod_{k < i \leq N}
(z_k - z_i)^2 \Psi(z_{k+1}, z_{k+2}, \dots, z_N).
\ee
Substituting \rf{theorem} to \rf{statenew} we see that
\be
\label{statex}
\VEV{ \psi_0 (z_k) \psi_1 (z_{k+1}) \dots \psi_1 (z_N) }
\propto
{\Psi (z_{k+1}, z_{k+2}, \dots, z_N) \over
\prod_{k <i < j \leq N}
(z_i-z_j)^{2 \over k}},
\ee
that is, 
$\psi_0$ is indeed an identity operator. The correlation function
is insensitive to its insertion. 
That completes the first part of our proof.

If we continue to expand \rf{statelast} in powers of $z_{k-1}-z_k$ we
observe that we should reproduce the operator product expansion
\rf{OPE} with $l+l' = 0$ mod $k$. In particular, we must see that
the linear term of that expansion vanishes, $\beta_{k-1,1}=0$. If this term
didn't vanish, that would mean that in addition to the parafermion fields,
the conformal field theory whose correlation function is given by
\rf{state} has a dimension 1 operator $J$ which generates a 
U$(1)$ symmetry (see \ct{ZF}). Such an operator should be absent in the
parafermion theory. Let us check that it is indeed absent. 

Continuing the expansion of \rf{statelast} in powers of $z_{k-1}-z_k$
we find that the linear term is obviously proportional to 
\be
\label{statelast1}
{\partial \over \partial z_{k-1}} \left. \left[
{\Psi (z_{k-1}, z_{k-1}, \dots, z_{k-1}, z_k, z_{k+1}, \dots, z_N) \over
\prod_{i>k} 
(z_{k-1}-z_i)^{2 (k-1) \over k}} \right] \right|_{z_{k-1}=z_k} .
\ee
To show that this term is zero it is sufficient to show that
\be
\label{theor}
\left. {\partial \over \partial z_{k-1}} 
\Psi(z_{k-1}, z_{k-1}, \dots, z_{k-1}, z_k, z_{k+1}, \dots, z_N) 
\right|_{z_{k-1}=z_k}
= {2 (k-1) \over k} \left( \sum_{i>k} {1 \over z_k-z_i} \right)
\Psi(z_{k}, z_{k}, \dots, z_{k}, z_{k+1}, \dots, z_N).
\ee
We can further simplify the equality \rf{theor} by noting that
since $\Psi$ is a symmetric function of its arguments, it is enough
to differentiate it with respect to just one variable,
\be
\label{theor1}
\left. {\partial \over \partial z_{k}} 
\Psi(z_{k-1}, z_{k-1}, \dots, z_{k-1}, z_k, z_{k+1}, \dots, z_N) 
\right|_{z_{k-1}=z_k}
= {2 \over k} \left( \sum_{i>k} {1 \over z_k-z_i} \right)
\Psi(z_{k}, z_{k}, \dots, z_{k}, z_{k+1}, \dots, z_N).
\ee
To prove \rf{theor1} 
we again recall the definition of $\Psi(z_1, \dots, z_N)$ as given in
\rf{wave}. To construct it, we break all its coordinates into
$N/k$ clusters of $k$ coordinates and then sum a certain expression,
written in terms of these clusters, over all the permutations
of $N$ particles. Let us first look at the right hand side of \rf{theor1}.
In accordance with the theorem proved in \ct{RR}, the only terms which
contribute to $\Psi(z_k, \dots, z_k, z_{k+1}, \dots, z_N)$ 
in the sum over permutations \rf{wave} are those
where all the $z_k$ belong to the same cluster. This is how
\rf{theorem} could be derived. Let us now make an assumption,
which we will justify later,
 that the
only terms which contribute to the left hand side of \rf{theor1} when 
we substitute \rf{wave} for $\Psi$ are also
those where all the $z_{k-1}$ and $z_k$ belong to the same cluster. 

It is possible to convince oneself that after some algebra the sum
of these terms can be reduced to 
\be
\Phi = \sum_{r=1}^k \sum_{P}
\prod_{0<j < N/k} {(z_k-z_{P\left(r+kj\right)}) (z_k-
z_{P\left(r+kj+s\left(r\right) \right)})
\over (z_{k-1}-z_{P\left(r+kj\right)})
(z_{k-1}-z_{P\left(r+kj+s\left(r\right) \right)})}   
\prod_{k<i \le N} (z_{k-1}-z_i)^2 
\prod_{0<j<l < N/k} \chi_{j+1, l+1} , 
\ee
where $P(n)$ gives a permutation of the integer numbers $k+1$, $k+2$, $\dots$,
$N$. And $s(n) = n+1$ for $0 < n < k$ 
with $s(k)=1$. 
$\chi$ are the expressions \rf{chi}. Note that $\chi$ 
do not depend on $z_{k-1}$ and $z_{k}$.  
Applying the derivatives as in \rf{theor1} we get
\bea
\label{permu}
\left. \pp{z_k} \Phi \right|_{z_{k-1}=z_k} = \br \sum_{r=1}^k \sum_{P}
\sum_{0<j < N/k} \left( {1 \over z_k-z_{P\left(r+kj\right)}} +
{1 \over z_k-z_{P\left(r+ k j+s\left(r\right) \right)}}
\right) \prod_{k<i \le N} (z_{k}-z_i)^2 
\prod_{0<j<l < N/k} \chi_{j+1,l+1} = \br 
{2 \over k}
\left(\sum_{k<i\le N} {1 \over z_k - z_i} \right) 
\left. \Phi \right|_{z_{k-1}=z_k} .
\eea
The last line in \rf{permu} follows from the fact that the summation over
$r$
completely symmetrizes
the sum in the second line of \rf{permu} over all the permutations
$P$.

Therefore we have proved \rf{theor1} on the assumptions that the only 
terms
that contribute
are those where the first $k$ coordinates of $\Psi$ belong to the same
cluster. To see that it is indeed so, let us recall
again that according to the theorem proved in \ct{RR} all such term should
vanish as $z_{k-1}$ approach $z_k$. Since they are polynomials they
vanish at least as $z_{k}-z_{k-1}$, or perhaps even faster. If they vanish
faster, their derivative with
respect to $z_k$ with the setting $z_{k-1}=z_k$ after differentiation
is definitely zero.
If, on the other hand, they vanish linearly, then we can argue that
for each term vanishing as $z_{k}-z_{k-1}$ there is another permutation
of the coordinates different from the first one by exchanging exactly
the two coordinates in that difference. This term will be proportional to
$z_{k-1}-z_k$. After differentiating  
{\sl and setting} $z_k=z_{k-1}$, these
terms will cancel each other.

This concludes the proof that the wave function conjectured in \ct{RR} to
be the correlation function of the parafermions is indeed the correlation
function of the parafermions. 
This in fact allows us to use the 
parafermionic conformal field theory to describe the ground states
of the hamiltonian \rf{Hamel}, as we did throughout this paper.

\section {Acknowledgements}

The authors are grateful to N. Read for important discussions and
to K. Schoutens for explaining the results of his paper
\ct{BS}. This work was initiated and completed during the
ITP program ``Disorder and Interaction in Quantum Hall and Mesoscopic
Systems''
and was supported by the NSF grants PHY-94-07194 and DMR-9420560 (ER).
ER is also grateful to ITP for an ITP Scholar award. 

\appendix
\section{Numerical Data}
In this appendix we present the numerical data 
for the degeneracy of the angular momenta multiplets at various 
excess flux quanta $n$.  Columns of the tables are labeled
by $n$. Rows are labeled by the
angular momentum $L$. The intersection of the $n$th column and $L$th row
gives us the degeneracy of the angular momentum $L$ multiplets at a given
$n$. The three tables represent the angular momentum multiplets at the
total particle number $N=6$, $N=9$, and $n=12$. All the data is at $k=3$.
\narrowtext
\begin{table}[htbp]
\begin{tabular}{c||c|c|c|c|c c c}
 $L \backslash n$ & 1 & 2 & 3 & 4 & 5 & & \\ \hline \hline
       0          &   & 2 &   & 3 &   & &\\ \hline
       1          & 1 &   & 2 &   & 4 & &\\ \hline
       2          &   & 2 & 1 & 4 & 2 & &\\ \hline
       3          & 1 & 1 & 4 & 3 & 7 & &\\ \hline
       4          &   & 2 & 2 & 6 & 5 & &\\ \hline
       5          &   &   & 3 & 3 & 8 & &\\ \hline
       6          &   & 1 & 2 & 6 & 7 & &\\ \hline
       7          &   &   & 2 & 3 & 8 & &\\ \hline
       8          &   &   &   & 4 & 5 & &\\ \hline
       9          &   &   & 1 & 2 & 7 & &\\ \hline
       10         &   &   &   & 2 & 4 & &\\ \hline
       11         &   &   &   &   & 4 & &\\ \hline
       12         &   &   &   & 1 & 2 & &\\ \hline
       13         &   &   &   &   & 2 & &\\ \hline
       14         &   &   &   &   &   & &\\ \hline
       15         &   &   &   &   & 1 & &\\ 
\end{tabular} 
\caption{$N=6.$}
\label{kabs}
\end{table}

\begin{table}[htbp]
\begin{tabular}{c||c|c|c|c c c c}
 $L \backslash n$ & 1 & 2 & 3 & 4 &   & & \\ \hline \hline
       0          &   &   &   & 6 &   & &\\ \hline
       1/2        & 1 &   & 2 &   &   & &\\ \hline
       1          &   & 3 &   & 5 &   & &\\ \hline
       3/2        & 1 &   & 6 &   &   & &\\ \hline
       2          &   & 1 &   & 14&   & &\\ \hline
       5/2        & 1 &   & 7 &   &   & &\\ \hline
       3          &   & 5 &   & 14&   & &\\ \hline
       7/2        &   &   & 8 &   &   & &\\ \hline
       4          &   & 2 &   & 21&   & &\\ \hline
       9/2        & 1 &   & 9 &   &   & &\\ \hline
       5          &   & 3 &   &17 &   & &\\ \hline
       11/2       &   &   & 9 &   &   & &\\ \hline
       6          &   & 2 &   &23 &   & &\\ \hline
       13/2       &   &   & 7 &   &   & &\\ \hline
       7          &   & 2 &   &18 &   & &\\ \hline
       15/2       &   &   & 8 &   &   & &\\ \hline
       8          &   &   &   &20 &   & &\\ \hline
       17/2       &   &   & 5 &   &   & &\\ \hline
       9          &   & 1 &   &16 &   & &\\ \hline
       19/2       &   &   & 4 &   &   & &\\ \hline
       10         &   &   &   &16 &   & &\\ \hline
       21/2       &   &   & 3 &   &   & &\\ \hline
       11         &   &   &   &10 &   & &\\ \hline
       23/2       &   &   & 2 &   &   & &\\ \hline
       12         &   &   &   &11 &   & &\\ \hline
       25/2       &   &   & 0 &   &   & &\\ \hline
       13         &   &   &   & 6 &   & &\\ \hline
       27/2       &   &   & 1 &   &   & &\\ \hline
       14         &   &   &   & 5 &   & &\\ \hline
       29/2       &   &   &   &   &   & &\\ \hline
       15         &   &   &   & 3 &   & &\\ \hline
       31/2       &   &   &   &   &   & &\\ \hline
       16         &   &   &   & 2 &   & &\\ \hline
       33/2       &   &   &   &   &   & &\\ \hline
       17         &   &   &   &   &   & &\\ \hline
       35/2       &   &   &   &   &   & &\\ \hline
       18         &   &   &   & 1 &   & &\\ 
\end{tabular} 
\caption{$N=9.$}
\label{nidra}
\end{table}

\begin{table}[htbp]
\begin{tabular}{c||c|c|ccccc}
 $L \backslash n$ & 1 & 2 & 3 &  &  & & \\ \hline \hline
       0          & 1 & 4 & 6 &  &   & &\\ \hline
       1          &   & 1 & 7 &   &  & &\\ \hline
       2          & 1 & 6 & 16&  &  & &\\ \hline
       3          & 1 & 4 & 18 &  &  & &\\ \hline
       4          & 1 & 8 & 24&  &  & &\\ \hline
       5          &   & 4 & 21&  &  & &\\ \hline
       6          & 1 & 7 & 27&  &  & &\\ \hline
       7          &   & 3 & 22&  &  & &\\ \hline
       8          &   & 4 & 22&  &  & &\\ \hline
       9          &   & 2 & 18&  &  & &\\ \hline
       10         &   & 2 & 17&  &  & &\\ \hline
       11         &   &   & 11&   &  & &\\ \hline
       12         &   & 1 & 11&  &  & &\\ \hline
       13         &   &   & 6 &   &  & &\\ \hline
       14         &   &   & 5 &   &   & &\\ \hline
       15         &   &   & 3 &   &  & &\\  \hline
       16         &   &   & 2 &   &   & &\\ \hline
       17         &   &   &   &   &   & &\\ \hline
       18         &   &   & 1 &   &   & &\\ 
\end{tabular} 
\caption{$N=12.$}
\label{nidraban}
\end{table}

\widetext

\end{document}